\documentclass[preprintnumbers]{revtex4}
\UseRawInputEncoding
\usepackage{amssymb}
\usepackage{amsmath}
\usepackage{graphicx}
\usepackage{dcolumn}
\usepackage{bm}
\usepackage{subfigure}
\usepackage{color}
\usepackage{multirow}

\begin{document}

\title{Thermodynamic relation on black branes with arbitrary cosmological constant}
\author{Yubo Ma$^{1*,2}$, Songtao Zheng$^{1}$, Yang Zhang$^{1,2}$, Mengsen Ma$^{1,2}$, Yunzhi Du$^{1*,2}$}
\affiliation{$^{1}$ Department of physics, Shanxi Datong University, Datong,037009,China\\ $^{2}$ Institute of Theoretical Physics, Shanxi Datong University, Datong,037009,China}

\thanks{\emph{e-mail:yuboma.phy@gmail.com,duyzh22@sxdtdx.edu.cn}}

\begin{abstract}

We investigated the Goon-Penco (GP) relationship in $(n+1)$-dimensional black branes with an arbitrary cosmological constant. Our analysis revealed that the GP relation preserved its form in four-dimensional and $(n+1)$-dimensional spacetimes, demonstrating its universal behavior with respect to dimensionality. Furthermore, we established that the GP relation exhibits universality across all states of the black hole, including those associated with the event horizon and the cosmological horizon. These findings confirm that the GP relationship remains valid for $(n+1)$-dimensional black holes and black branes with an arbitrary cosmological constant, independent of the coordinate system employed.
\end{abstract}

\maketitle
\section{Introduction}
\label{sec:intro}
The Weak Gravity Conjecture (WGC) postulates that in any consistent theory of quantum gravity, the gravitational force must be weaker than the force exerted by some lighter, charged particle or field \cite{Vafa05,Arkani-Hamed07}. Specifically, it reveals that there exists a particle with charge $Q$ and mass $M$ such that the force it experiences due to gravity is weaker than the force it experiences due to its gauge interaction i.e., $Q<M$. However, when the extremal black hole (BH) saturates the bound for a charged BH, the relation of the charge $Q$ and the mass $M$ is written as $Q\le M$. This conjecture is especially important in the context where global symmetries are absent \cite{Banks11} and is based on the principles of quantum gravity \cite{Harlow23}. Building on studies on the WGC, Goon and Penco investigated the universality of the thermodynamic relation between entropy and extremality under perturbation \cite{Goon20}. Perturbations in free energy yield a relationship between mass, temperature and entropy with corrections. The leading-order expansion of perturbative parameters reveals an approximate relation that can be linked to higher-derivative corrections \cite{Wang23}, generating a connection between shifts in entropy and the charge-to-mass ratio \cite{Cheung19,Kats07,Cheung18}. Based on the relation proposed by Goon and Penco, considerable progress has been made, including analysis of the relation between various AdS spacetimes such as charged BTZ black holes and Kerr-AdS black holes from the WGC perspective \cite{Cano20a,Sadeghi23,Cremonini20,Cano20b,Alipour24,Ko25,Cremonini23,Lee24,Ma23a,Wei21,Reall19,Alipour23,Ma21}.
It is well known that for dS spacetime, when certain conditions are met in the spacetime parameters, there exists a coexistence region with both a black hole event horizon and a cosmological event horizon \cite{McPeak22,Sadeghi22}. In this region, the mass $M$ of spacetime as a function of $r$exhibits two extrema, with corresponding masses denoted as ${{M}_{N}}$ and ${{M}_{C}}$. Refs. \cite{Nam18,Zhen24} apply the theory from Ref.\cite{Banks11} to study the relationship $M\to {{M}_{N}}$ in Reissner-Nordstr$\ddot{o}$m-de Sitter (RNdS), Kerr-de Sitter (KdS), and Kerr-Newman-de Sitter (KNdS) spacetimes. The universality of the thermodynamic relation is confirmed in dS black holes. Furthermore, the results suggest that the Weak Gravity Conjecture (WGC) is still applicable in dS spacetime. However, up until now, the models studied have mostly used spherical coordinates to describe black hole spacetimes. There have been no reports on models describing complex spacetimes in cylindrical coordinates. This paper will apply the viewpoint from Ref.\cite{Goon20} to discuss the universal thermodynamic relations for $n$-dimensional black holes described in cylindrical coordinates. The results show that the WGC is still applicable to spacetimes described in cylindrical coordinates, thus proving the universality of the thermodynamic relation.
The structure of the paper is as follows: In Section 2, we combine the first law of thermodynamics in dS spacetime to discuss the universal relation of the Einstein-Maxwell-Dilaton action in $(n+1)$-dimensional spacetime$(n\ge 3)$. We derive the relationship between the thermodynamic quantities corresponding to the black hole event horizon and the cosmological event horizon as a function of the perturbation parameter $\eta $, under the condition that the energy $M$ of spacetime satisfies the coexistence of both the black hole horizon and the cosmological horizon.Section 3 discusses the universal relations for the $(n+1)$-dimensional Einstein-Maxwell-Dilaton gravity spacetime, providing the universal relation for the black brane when both charge $Q$ and $J$ are functions of the perturbation parameter $\eta $. We prove that the relation given by Goon and Penco is independent of the choice of coordinate system and is universal for spacetimes with event horizons.Section 4 provides the conclusion.

\section{Review of Topological Charged Dilaton Black Holes in de Sitter Space}
\label{sec:method}
The Einstein-Maxwell-Dilaton action in $(n + 1)$-dimensional $(n \ge 3)$spacetime is \cite{Ko24,Ma25,Dehghani14}
\begin{equation}\label{2.1}
S = \frac{1}{{16\pi }}\int {{d^{n + 1}}} x\sqrt { - g} \left( {R - \frac{4}{{n - 1}}{{(\nabla \Phi )}^2} - V(\Phi ) - {e^{ - 4\alpha \Phi /(n - 1)}}{F_{\mu \nu }}{F^{\mu \nu }}} \right)
\end{equation}
where $R$ is the Ricci scalar curvature, the dilaton potential $V\left( \Phi  \right)$ is expressed in terms of the dilaton field $\Phi$ and its coupling to the cosmological constant $\Lambda$,
\begin{equation}\label{2.2}
\begin{aligned}
&V\left( \Phi  \right) = 2\Lambda {e^{{{4\alpha \Phi } \mathord{\left/
 {\vphantom {{4\alpha \Phi } {\left( {n - 1} \right)}}} \right. \kern-\nulldelimiterspace} {\left( {n - 1} \right)}}}} + \frac{{k\left( {n - 1} \right)\left( {n - 2} \right){\alpha ^2}}}{{{b^2}\left( {{\alpha ^2} - 1} \right)}}{e^{4{\Phi  \mathord{\left/
 {\vphantom {\Phi  {\left[ {\left( {n - 1} \right)\alpha } \right]}}} \right.
 \kern-\nulldelimiterspace} {\left[ {\left( {n - 1} \right)\alpha } \right]}}}},\\
 &{\nabla ^2}\Phi  = \frac{{n - 1}}{8}\frac{{\partial V}}{{\partial \Phi }} - \frac{\alpha }{2}{e^{ - 4\alpha \Phi /(n - 1)}}{F_{\lambda \eta }}{F^{\lambda \eta }}.
 \end{aligned}
\end{equation}

\begin{equation}\label{2.3}
{\partial _\mu }\left( {\sqrt { - g} {e^{ - 4\alpha \Phi /(n - 1)}}{F^{\mu \nu }}} \right) = 0
\end{equation}
$\alpha $ is a constant determining the strength of coupling of the scalar and electromagnetic field, ${{F}_{\mu \nu }}={{\partial }_{\mu }}{{A}_{\nu }}-{{\partial }_{\nu }}{{A}_{\mu }}$ is the electromagnetic field tensor and ${{A}_{\mu }}$ is the electromagnetic vector potential.

The topological black hole solutions take the form Ref.\cite{Dayyani17}

\begin{equation}\label{2.4}
d{{s}^{2}}=-f(r)d{{t}^{2}}+\frac{d{{r}^{2}}}{f(r)}+{{r}^{2}}{{R}^{2}}(r)d\Omega _{n-1}^{2}
\end{equation}

where
\begin{equation}\label{2.5}
\begin{aligned}
f(r)&=\frac{(n-2){{({{\alpha }^{2}}+1)}^{2}}{{b}^{-2\gamma }}{{r}^{2\gamma }}}{(1-{{\alpha }^{2}})({{\alpha }^{2}}+n-2)}-\frac{16\pi ({{\alpha }^{2}}+1)M}{{{b}^{(n-1)\gamma }}(n-1){{\omega }_{n-1}}{{r}^{(n-1)(1-\gamma )-1}}}\\
&+\frac{32{{\pi }^{2}}{{Q}^{2}}{{({{\alpha }^{2}}+1)}^{2}}{{b}^{-2(n-2)\gamma }}}{\omega _{n-1}^{2}(n-1)({{\alpha }^{2}}+n-2)}{{r}^{2(n-2)(\gamma -1)}}-\frac{n(1+\eta ){{({{\alpha }^{2}}+1)}^{2}}{{b}^{2\gamma }}}{{{l}^{2}}(n-{{\alpha }^{2}})}{{r}^{2(1-\gamma )}}\\
&=-A{{r}^{2\gamma }}-\frac{BM}{{{r}^{(n-1)(1-\gamma )-1}}}+{{Q}^{2}}{{r}^{2(n-2)(\gamma -1)}}C-\frac{(1+\eta )D}{{{l}^{2}}}{{r}^{2(1-\gamma )}},
\end{aligned}
\end{equation}

\begin{equation}\label{2.6}
R(r)={{e}^{2\alpha \Phi \left( r \right)/(n-1)}},\Phi (r)=\frac{(n-1)\alpha }{2(1+{{\alpha }^{2}})}\ln \left( \frac{b}{r} \right),
\end{equation}

with
\begin{equation}\label{2.7}
\begin{aligned}
&A=\frac{(n-2){{({{\alpha }^{2}}+1)}^{2}}{{b}^{-2\gamma }}}{({{\alpha }^{2}}-1)({{\alpha }^{2}}+n-2)}, B=\frac{16\pi ({{\alpha }^{2}}+1)}{{{b}^{(n-1)\gamma }}(n-1){{\omega }_{n-1}}},\\
&C=\frac{32{{\pi }^{2}}{{({{\alpha }^{2}}+1)}^{2}}{{b}^{-2(n-2)\gamma }}}{\omega _{n-1}^{2}(n-1)({{\alpha }^{2}}+n-2)},  D=\frac{n{{({{\alpha }^{2}}+1)}^{2}}{{b}^{2\gamma }}}{(n-{{\alpha }^{2}})}.
\end{aligned}
\end{equation}
in which $\gamma ={{\alpha }^{2}}/({{\alpha }^{2}}+1)$, $b$ is an arbitrary constant, $l$ denotes the ADS length scale, the cosmological constant $\Lambda$ is related to spacetime dimension $n$ and $l$ by
\begin{equation}\label{2.8}
\Lambda =\frac{n(1+\eta )(n-1)}{2{{l}^{2}}},
\end{equation}
$M$ is the ADM (Arnowitt-Deser-Misnsr) mass of the black hole related to an integration constant $m$ \cite{Dayyani17,Zhao21} and the electric charge is related to the parameter $q$,

\begin{equation}\label{2.9}
M=\frac{{{b}^{(n-1)\gamma }}(n-1){{\omega }_{n-1}}}{16\pi ({{\alpha }^{2}}+1)}m,  Q=\frac{q{{\omega }_{n-1}}}{4\pi },
\end{equation}
where ${{\omega }_{n-1}}$ represents the volume of constant curvature hypersurface described by $d\Omega _{k,n-1}^{2}$. When $\Lambda >0$, the dS spacetime has black hole event horizon (BEH) and cosmological event horizon (CEH), which locate respectively at $r={{r}_{+}}$ and $r={{r}_{c}}$. They meet $f({{r}_{+}})=0$ and $f({{r}_{c}})=0$.

The Hawking radiation temperatures ${{T}_{+}}$ and ${{T}_{c}}$ on BEH and CEH respectively satisfy

\begin{equation}\label{2.10}
\begin{aligned}
2\pi {{T}_{+}}&={{\kappa }_{+}}=\frac{1}{2}{{\left. \frac{df(r)}{dr} \right|}_{r={{r}_{+}}}}=\frac{({{\alpha }^{2}}+1)}{(n-1)}\\
&\left( \frac{(n-2)(n-1){{b}^{-2\gamma }}}{2(1-{{\alpha }^{2}})}r_{+}^{2\gamma -1}-\frac{n(1+\eta )(n-1)}{2{{l}^{2}}}{{b}^{2\gamma }}r_{+}^{1-2\gamma }-{{q}^{2}}{{b}^{-2(n-2)\gamma }}r_{+}^{(2n-3)(\gamma -1)-\gamma } \right),
\end{aligned}
\end{equation}
and
\begin{equation}\label{2.11}
\begin{aligned}
&2\pi {{T}_{c}}={{\kappa }_{c}}=-\frac{1}{2}{{\left. \frac{df(r)}{dr} \right|}_{r={{r}_{c}}}}=-\frac{({{\alpha }^{2}}+1)}{(n-1)}\\
&\left( \frac{(n-2)(n-1){{b}^{-2\gamma }}}{2(1-{{\alpha }^{2}})}r_{c}^{2\gamma -1}-\frac{n(1+\eta )(n-1)}{2{{l}^{2}}}{{b}^{2\gamma }}r_{c}^{1-2\gamma }-\frac{16{{\pi }^{2}}{{Q}^{2}}}{\omega _{n-1}^{2}}{{b}^{-2(n-2)\gamma }}r_{c}^{(2n-3)(\gamma -1)-\gamma } \right),
\end{aligned}
\end{equation}

\begin{equation}\label{2.12}
M=\frac{1}{B}r_{+,c}^{(n-1)(1-\gamma )-1}\left( -Ar_{+,c}^{2\gamma }+{{Q}^{2}}r_{+,c}^{2(n-2)(\gamma -1)}C-\frac{(1+\eta )D}{{{l}^{2}}}r_{+,c}^{2(1-\gamma )} \right),
\end{equation}
\begin{equation}\label{2.13}
\frac{\partial M}{\partial \eta }=\frac{1}{B}r_{+,c}^{(n-1)(1-\gamma )-1}\left( -Ar_{+,c}^{2\gamma }+{{Q}^{2}}r_{+,c}^{2(n-2)(\gamma -1)}C-\frac{D}{{{l}^{2}}}r_{+,c}^{2(1-\gamma )} \right).
\end{equation}
When $\alpha $, $b$, $Q$ and the spacetime dimension $n$are given, and with $M$ remaining constant (i.e.$M={{M}_{NC}}$, ${{M}_{c}}\le {{M}_{NC}}\le {{M}_{N}}$,where ${{M}_{N}}$ represents the highest energy corresponding to the coexistence region of two horizons, and ${{M}_{C}}$ represents the lowest energy corresponding to the coexistence region), from Eq.(\ref{2.12}), we know that when $M={{M}_{NC}}$, as the perturbation parameter $\eta $ changes, the position of the event horizon ${{r}_{+,c}}$changes accordingly due to the constraint of Eq.(\ref{2.12})(i.e., the entropy changes). From this, we obtain

\begin{equation}\label{2.14}
\begin{aligned}
0=&\frac{\partial }{\partial {{S}_{+,c}}}\left[ \frac{1}{B}r_{+,c}^{(n-1)(1-\gamma )-1}\left( -Ar_{+,c}^{2\gamma }+{{Q}^{2}}r_{+,c}^{2(n-2)(\gamma -1)}C-\frac{(1+\eta )D}{{{l}^{2}}}r_{+,c}^{2(1-\gamma )} \right) \right]\frac{\partial {{S}_{+,c}}}{\partial \eta }\\
&+\frac{1}{B}r_{+,c}^{(n-1)(1-\gamma )-1}\left( -Ar_{+,c}^{2\gamma }+{{Q}^{2}}r_{+,c}^{2(n-2)(\gamma -1)}C-\frac{D}{{{l}^{2}}}r_{+,c}^{2(1-\gamma )} \right)
\end{aligned}
\end{equation}

By comparing Eqs. (\ref{2.13}) and (\ref{2.14}), we get
\begin{equation}\label{2.15}
\begin{aligned}
\frac{\partial M}{\partial {{S}_{+,c}}}\frac{\partial {{S}_{+,c}}}{\partial \eta }&=-\frac{1}{B}r_{+,c}^{(n-1)(1-\gamma )-1}\left( -Ar_{+,c}^{2\gamma }+{{Q}^{2}}r_{+,c}^{2(n-2)(\gamma -1)}C-\frac{D}{{{l}^{2}}}r_{+,c}^{2(1-\gamma )} \right)=-\frac{\partial M}{\partial \eta },\\
&\frac{\partial {{M}_{NC}}}{\partial \eta }=\mp \underset{M\to {{M}_{NC}}}{\mathop{\lim }}\,{{T}_{+,c}}{{\left( \frac{\partial {{S}_{+,c}}}{\partial \eta } \right)}_{a,b,Q,M}}
\end{aligned}
\end{equation}

When the thermodynamic quantities corresponding to the black hole event horizon and the cosmological event horizon satisfy the first law of thermodynamics \cite{Sheykhi07,Abbott82,Mbarek19,Simovic19,Simovic18,Dolan13}, we can use a simple method to derive the conclusion of the Goon and Penco relation in Eq.(\ref{2.15}).
When $\alpha $, $b$, $Q$ and the spacetime dimension $n$ are given, under the constraint of Eq.(\ref{2.12}), the change in energy due to the variation of the perturbation parameter $\eta $ is

\begin{equation}\label{2.16}
\begin{aligned}
\frac{dM}{d\eta }&=\left( \frac{\partial M}{\partial \eta } \right)+\left( \frac{\partial M}{\partial {{r}_{+,c}}} \right)\left( \frac{\partial {{r}_{+,c}}}{\partial \eta } \right)\\
&=\left( \frac{\partial M}{\partial \eta } \right)+\left( \frac{\partial M}{\partial {{S}_{+,c}}} \right)\left( \frac{\partial {{S}_{+,c}}}{\partial \eta } \right)=\left( \frac{\partial M}{\partial \eta } \right)\pm {{T}_{+,c}}\left( \frac{\partial {{S}_{+,c}}}{\partial \eta } \right).
\end{aligned}
\end{equation}

When the energy $M={{M}_{NC}}$ is given, Eq.(\ref{2.16}) becomes
\begin{equation}\label{2.17}
{{\left( \frac{\partial {{M}_{NC}}}{\partial \eta } \right)}_{Q,a,b}}=\mp \underset{M\to {{M}_{NC}}}{\mathop{\lim }}\,{{T}_{+,c}}{{\left( \frac{\partial S}{\partial \eta } \right)}_{Q,a,b,M}}.                \end{equation}

The given conclusion is independent of the spacetime dimension.

\section{The $(n + 1)$-dimensional Einstein-Maxwell-dilaton gravity}
The metric of $(n + 1)$-dimensional rotating solution with cylindrical or toroidal horizons and  rotation parameters can be written as \cite{Cai02,Sekiwa06,Awad03}

\begin{equation}\label{3.1}
\begin{aligned}
d{s^2} = & - f(r){\left( {\Xi dt - \sum\limits_{i = 1}^N {{a_i}d{\phi _i}} } \right)^2} + {r^2}(1 + \eta )g{R^2}(r)\sum\limits_{i = 1}^N {((1 + \eta )g{a_i}} dt - \Xi d{\phi _i}{)^2}\\
&- {r^2}(1 + \eta )g{R^2}(r)\sum\limits_{i < j}^N {({a_i}} d{\phi _j} - {a_j}d{\phi _i}{)^2} + \frac{{d{r^2}}}{{f(r)}} + {r^2}(1 + \eta )g{R^2}(r)d{X^2},
\end{aligned}
\end{equation}
here
\begin{equation}\label{3.2}
 {\Xi ^2} = 1 + \sum\limits_{i = 1}^N {(1 + \eta )ga_i^2},
 \end{equation}
 ${{a}_{i}}^{}$ and $k$ are rotation parameters. The function $f(r)$ and $R(r)$ should be determined and $g$ has the dimension of length which is related to the cosmological constant $\Lambda $ by the relation $2(1+\eta )\Lambda =-n(n-1)g$. The angular coordinates are in the range $0\le {{\phi }_{i}}\le 2\pi $ and $d{{X}^{2}}$is the Euclidean metric on the $(n-k-1)$-dimensional sub-manifold with volume $\sum\nolimits_{n-k-1}{{}}$. By solving, we obtain

 \begin{equation}\label{3.3}
 f(r) = \frac{{2\Lambda {{({\alpha ^2} + 1)}^2}{b^{2\gamma }}{r^{2(1 - \gamma )}}}}{{(n - 1)({\alpha ^2} - n)}} - \frac{m}{{{r^{(n - 1)(1 - \gamma ) - 1}}}} + \frac{{2{q^2}{{({\alpha ^2} + 1)}^2}{b^{ - 2(n - 2)\gamma }}{r^{2(n - 2)(1 - \gamma )}}}}{{(n - 1)(({\alpha ^2} + n - 2)}},
 \end{equation}
 The thermodynamic volume $V$, entropy $S$, and effective potential ${\varphi _{eff}}$ of the system are given by the following expressions:

 \begin{equation}\label{3.4}
 \begin{aligned}
  \Phi (r) = \frac{{(n - 1)\alpha }}{{2(1 + {\alpha ^2})}}\ln \left( {\frac{b}{r}} \right),
\end{aligned}
  \end{equation}
where $b$ and $m$ are arbitrary constants and $\gamma ={{\alpha }^{2}}({{\alpha }^{2}}+1)$. Denoting the volume of the hypersurface boundary at constant $t$ and $r$ by ${{V}_{n-1}}={{(2\pi )}^{k}}\sum\nolimits_{n-k-1}{{}}$, the mass and angular momentum per unit volume ${{V}_{n-1}}$ of the black branes $(\alpha <\sqrt{n})$.

\begin{equation}\label{3.5}
M=\frac{{{b}^{(n-1)\gamma }}}{16\pi }{{[(1+\eta )g]}^{(n-2)/2}}\left( \frac{(n-{{\alpha }^{2}}){{\Xi }^{2}}+{{\alpha }^{2}}-1}{1+{{\alpha }^{2}}} \right)m.
\end{equation}

\begin{equation}\label{3.6}
{{J}_{i}}=\frac{{{b}^{(n-1)\gamma }}}{16\pi }{{[(1+\eta )g]}^{(n-2)/2}}\left( \frac{(n-{{\alpha }^{2}})}{1+{{\alpha }^{2}}} \right)\Xi m{{a}_{i}}.
\end{equation}

The entropy per volume ${{V}_{n-1}}$ of the black brane is
\begin{equation}\label{3.7}
\begin{aligned}
&S=\frac{\Xi {{b}^{(n-1)\gamma }}r_{+}^{(n-1)(1-\gamma )}}{4}{{[(1+\eta )g]}^{(n-2)/2}},Q=\frac{\Xi q}{4\pi }{{[(1+\eta )g]}^{(n-2)/2}},\\
&U=\frac{q{{b}^{(1-n)\gamma }}}{\Xi \Gamma r_{+,c}^{\Gamma }}{{[(1+\eta )g]}^{(n-2)/2}},{{\Omega }_{i}}=\frac{{{a}_{i}}}{\Xi }(1+\eta )g.
\end{aligned}
\end{equation}
where $\Gamma =(n-3)(1-\gamma )+1$.
\begin{equation}\label{3.8}
M = (S,J,Q) = {[(1 + \eta )g]^{1/2}}\frac{{[(n - {\alpha ^2}){\Xi ^2} + {\alpha ^2} - 1]J}}{{\Xi \sqrt {{\Xi ^2} - 1} }}
\end{equation}

\begin{equation}\label{3.9}
{{T}_{+}}=\frac{1}{4\pi \Xi }\left( \frac{(n-{{\alpha }^{2}})m}{{{\alpha }^{2}}+1}r_{+}^{(n-1)(\gamma -1)}-\frac{4{{q}^{2}}({{\alpha }^{2}}+1){{b}^{-2(n-2)\gamma }}}{({{\alpha }^{2}}+n-2)}r_{+}^{(2n-3)(\gamma -1)} \right).
\end{equation}

In the equation, ${{r}_{+}}$ satisfies $f({{r}_{+}})=0$,${{J}^{2}}=\sum\nolimits_{i}^{k}{J_{i}^{2}}$and ${{\Xi }^{2}}$ is the positive real root of the following equation
\begin{equation}\label{3.10}
\begin{aligned}
&(n-1)({{\alpha }^{2}}+n-2)\\
&\left[ n({{\alpha }^{2}}+1){{((1+\eta )g)}^{(n-2)/2}}{{\Xi }^{3}}\sqrt{{{\Xi }^{2}}-1}{{\left( \frac{4S}{{{\Xi }^{2}}{{((1+\eta )g)}^{(n-2)/2}}} \right)}^{(n-{{\alpha }^{2}})/(n-1)}}-\frac{16{{\Xi }^{2}}\pi {{b}^{-{{\alpha }^{2}}}}}{{{((1+\eta )g)}^{1/2}}} \right]\\
&=-\frac{32\Xi \sqrt{{{\Xi }^{2}}-1}}{{{((1+\eta )g)}^{n/2}}}{{\left( \frac{4S}{{{\Xi }^{2}}{{((1+\eta )g)}^{(n-2)/2}}} \right)}^{(n+{{\alpha }^{2}}-2)/(1-n)}}(n-{{\alpha }^{2}})({{\alpha }^{2}}+1){{\pi }^{2}}{{Q}^{2}}
\end{aligned}
\end{equation}

One may then regard the parameters $S$, $J$ and $Q$ as a complete set of extensive parameters for the mass $M(S,J,Q)$ and define the intensive parameters conjugate to $S$, $J$ and $Q$ These quantities are the temperature, the angular velocities and the electric potential

\begin{equation}\label{3.11}
{{T}_{+}}={{\left( \frac{\partial M}{\partial S} \right)}_{J,Q}}. {{\Omega }_{i}}={{\left( \frac{\partial M}{\partial {{J}_{i}}} \right)}_{S,Q}}, U={{\left( \frac{\partial M}{\partial Q} \right)}_{J,S}},
\end{equation}
these thermodynamics quantities satisfy the first law of thermodynamics \cite{Sheykhi06}
\begin{equation}\label{3.12}
dM={{T}_{+}}dS+\sum\limits_{i=1}^{k}{{{\Omega }_{i}}}d{{J}_{i}}+UdQ.
\end{equation}

From Eqs.(\ref{3.6}) and (\ref{3.7}), we know that when $g$ remains constant, both ${{J}_{i}}$and $Q$ are functions of $\eta $. When the parameter$\eta $in spacetime undergoes a perturbation, the corresponding change in energy $M$ is given by Eq.(\ref{3.8}).

\begin{equation}\label{3.13}
\begin{aligned}
\frac{dM}{d\eta }&={{\left( \frac{\partial M}{\partial \eta } \right)}_{g,S,J,Q}}+{{\left( \frac{\partial M}{\partial S} \right)}_{g,J,Q}}\left( \frac{\partial S}{\partial \eta } \right)+\sum\limits_{i=1}^{k}{{{\left( \frac{\partial M}{\partial {{J}_{i}}} \right)}_{g,S,Q}}}\left( \frac{\partial {{J}_{i}}}{\partial \eta } \right)+{{\left( \frac{\partial M}{\partial Q} \right)}_{g,J,S}}\left( \frac{\partial Q}{\partial \eta } \right)\\
&={{\left( \frac{\partial M}{\partial \eta } \right)}_{g,S,J,Q}}+{{T}_{+}}\left( \frac{\partial S}{\partial \eta } \right)+\sum\limits_{i=1}^{k}{{{\Omega }_{i}}}\left( \frac{\partial {{J}_{i}}}{\partial \eta } \right)+U\left( \frac{\partial Q}{\partial \eta } \right).
\end{aligned}
\end{equation}

When $M$ is given a certain value ${{M}_{0}}$, under the constraint of Eq.(\ref{3.8}), we obtain that ${{M}_{0}}$ must satisfy the condition for the existence of the black brane. i.e.

\begin{equation}\label{3.14}
\begin{aligned}
{{\left( \frac{\partial {{M}_{0}}}{\partial \eta } \right)}_{S,{{J}_{i}}(i=1,2,\cdots k),Q}}=&-\underset{M\to {{M}_{0}}}{\mathop{\lim }}\,\left[ {{T}_{+}}{{\left( \frac{\partial S}{\partial \eta } \right)}_{g,{{J}_{i}}(i=1,2,\cdots k),M,Q}}\right.\\
&\left.+\sum\limits_{i=1}^{k}{{{\Omega }_{i}}}{{\left( \frac{\partial {{J}_{i}}}{\partial \eta } \right)}_{g,S,Q,{{J}_{j}}(i\ne j=1,2,\cdots k),M}}+U{{\left( \frac{\partial Q}{\partial \eta } \right)}_{g,S,{{J}_{i}}(i=1,2,\cdots k),M}} \right]
\end{aligned}
\end{equation}

When $M$ is given a certain value ${{M}_{0}}$, from the constraint of Eq.(\ref{3.8}), we know that when the parameter $\eta $ in spacetime undergoes a perturbation, at least one state variable in the parameter $(S(\eta ),{{J}_{i(i=1,2,\cdots k)}}(\eta ),Q(\eta ))$ on the right-hand side of Eq.(\ref{3.8}) must change as $\eta $ changes. Therefore, we can take different state variables to change as $\eta $ changes. That is, when $\eta $ undergoes a perturbation, with ${{J}_{i(i=1,2,\cdots k)}}(\eta )$ and $Q(\eta )$ remaining constant, Eq.(\ref{3.14}) becomes

\begin{equation}\label{3.15}
{{\left( \frac{\partial {{M}_{0}}}{\partial \eta } \right)}_{g,S,{{J}_{i}}(i=1,2,\cdots k),Q}}=-\underset{M\to {{M}_{0}}}{\mathop{\lim }}\,{{T}_{+}}{{\left( \frac{\partial S}{\partial \eta } \right)}_{g,{{J}_{i}}(i=1,2,\cdots k),M,Q}},
\end{equation}

When $\eta$ undergoes a perturbation, with $Q(\eta )$ remaining constant, Eq.(\ref{3.14}) becomes

\begin{equation}\label{3.16}
\begin{aligned}
&{{\left( \frac{\partial {{M}_{0}}}{\partial \eta } \right)}_{S,{{J}_{i}}(i=1,2,\cdots k),Q}}\\
&=-\underset{M\to {{M}_{0}}}{\mathop{\lim }}\,\left[ {{T}_{+}}{{\left( \frac{\partial S}{\partial \eta } \right)}_{g,{{J}_{i}}(i=1,2,\cdots k),M,Q}}+\sum\limits_{i=1}^{k}{{{\Omega }_{i}}}{{\left( \frac{\partial {{J}_{i}}}{\partial \eta } \right)}_{g,S,Q,{{J}_{j}}(i\ne j=1,2,\cdots k),M}} \right],
\end{aligned}
\end{equation}

When $\eta$ undergoes a perturbation, with ${{J}_{i(i=1,2,\cdots k)}}(\eta )$remaining constant, Eq.(\ref{3.14}) becomes
\begin{equation}\label{3.17}
\begin{aligned}
&{{\left( \frac{\partial {{M}_{0}}}{\partial \eta } \right)}_{S,{{J}_{i}}(i=1,2,\cdots k),Q}}\\
&=-\underset{M\to {{M}_{0}}}{\mathop{\lim }}\,\left[ {{T}_{+}}{{\left( \frac{\partial S}{\partial \eta } \right)}_{g,{{J}_{i}}(i=1,2,\cdots k),M,Q}}+U{{\left( \frac{\partial Q}{\partial \eta } \right)}_{g,S,{{J}_{i}}(i=1,2,\cdots k),M}} \right], (3.17)
\end{aligned}
\end{equation}

When $\eta $ undergoes a perturbation, with $S(\eta )$ remaining constant, Eq.(\ref{3.14}) becomes
\begin{equation}\label{3.18}
\begin{aligned}
&{{\left( \frac{\partial {{M}_{0}}}{\partial \eta } \right)}_{S,{{J}_{i}}(i=1,2,\cdots k),Q}}\\
&=-\underset{M\to {{M}_{0}}}{\mathop{\lim }}\,\left[ \sum\limits_{i=1}^{k}{{{\Omega }_{i}}}{{\left( \frac{\partial {{J}_{i}}}{\partial \eta } \right)}_{g,S,Q,{{J}_{j}}(i\ne j=1,2,\cdots k),M}}+U{{\left( \frac{\partial Q}{\partial \eta } \right)}_{g,S,{{J}_{i}}(i=1,2,\cdots k),M}} \right],
\end{aligned}
\end{equation}

When $\eta $ undergoes a perturbation, with ${{J}_{\beta }}(\eta )$ remaining constant, Eq.(\ref{3.14}) becomes
\begin{equation}\label{3.19}
\begin{aligned}
&{{\left( \frac{\partial {{M}_{0}}}{\partial \eta } \right)}_{S,{{J}_{i}}(i=1,2,\cdots k),Q}}\\
&=-\underset{M\to {{M}_{0}}}{\mathop{\lim }}\,\left[ {{T}_{+}}{{\left( \frac{\partial S}{\partial \eta } \right)}_{g,{{J}_{i}}(i=1,2,\cdots k),M,Q}} \right.+\sum\limits_{i=1}^{\beta -1}{{{\Omega }_{i}}}{{\left( \frac{\partial {{J}_{i}}}{\partial \eta } \right)}_{g,S,Q,{{J}_{j}}(i\ne j=1,2,\cdots \beta -1),M}}\\
&\left. +\sum\limits_{i=\beta +1}^{k}{{{\Omega }_{i}}}{{\left( \frac{\partial {{J}_{i}}}{\partial \eta } \right)}_{g,S,Q,{{J}_{j}}(i\ne j=\beta +1,\beta +2\cdots k),M}}+U{{\left( \frac{\partial Q}{\partial \eta } \right)}_{g,S,{{J}_{i}}(i=1,2,\cdots k),M}} \right]
\end{aligned}
\end{equation}

From this, we obtain that when the energy $M$ is a multi-parameter function of
$(S(\eta ),\\{{J}_{i(i=1,2,\cdots k)}}(\eta ),Q(\eta ))$, by choosing different invariants, we derive different thermodynamic expressions of the Goon and Penco relation.

\section{Conclusion}
In the above discussion, we found that when the other parameters of spacetime are fixed, and the perturbation parameter $\eta $ changes, the state variables and the positions of the event horizons vary accordingly with $\eta $. Using this conclusion, we verify the findings of Ref.\cite{Goon20} and extend their results to the coexistence region of two horizons, as well as generalize the study to arbitrary dimensions of complex spacetimes. This method is simple in computation and clear in process. It provides a pathway for further in-depth exploration of the physical properties of the coexistence region of two horizons in dS spacetime. By applying the approach from Ref.\cite{Goon20} to the description of black branes in cylindrical coordinates, we derive the multi-parameter Goon and Penco relation when the energy $M$ of spacetime is a multi-parameter $(S(\eta ),{J_{i(i = 1,2, \cdots k)}}(\eta ),Q(\eta ))$ function. By choosing different invariants, we obtain the multi-parameter Goon and Penco relation.The exploration of the proportional relationship between corrected mass and entropy provides a pathway to understanding the Weak Gravity Conjecture (WGC). The WGC offers intriguing insights into the realm of quantum gravity, and studies such as the one presented in this paper contribute to a deeper understanding of this fundamental aspect of physics.
In conclusion, our calculations reveal the universal Goon and Penco relation for black hole spacetimes with multi-parameters in different coordinate systems. The exploration of the proportional relationship between corrected mass and entropy provides a new pathway for understanding the WGC. Thus, the results of this study can facilitate a deeper understanding of quantum gravity. Further, these results are expected to provide additional motivation for future studies in this field.

\section*{Acknowledgments}
This work was supported by the National Natural Science Foundation of China a under Grant No. 12075143 and No. 12375050, the Natural Science Foundation of Shanxi Province, China under Grant Nos. 202203021221211, 202203021221209, 202303021211180, the Innovative Talents of Higher Education Institutions of Shanxi with Grant No. 2024Q030, and the Teaching Reform Project of Shanxi Datong Universtiy with Grant No. XJG2022234.

\end{document}